\documentclass[preprint,12pt]{elsarticle}

\usepackage{graphicx}
\usepackage{amssymb}
\usepackage{amsmath}

\usepackage{float}
\usepackage{multicol}
\usepackage{multirow}
\usepackage{tabularx}
\usepackage{soul}
\usepackage{color}

\usepackage{lineno}

\journal{Physical Review Fluids}

\begin{document}

\begin{frontmatter}

\title{Pressure scrambling effects and the quantification of turbulent scalar flux model uncertainties}

\author{Zengrong Hao, Catherine Gorl\'e}

\address{Wind Engineering Laboratory, Department of Civil and Environmental Engineering, Stanford University, Stanford, CA 94305, USA}

\begin{abstract}

Closure models for the turbulent scalar flux are an important source of uncertainty in Reynolds-averaged-Navier-Stokes (RANS) simulations of scalar transport. This paper presents an approach to quantify this uncertainty in simulations of complex engineering flows. The approach addresses the uncertainty in modeling the pressure scrambling (PS) effect, which is the primary mechanism balancing the productions in scalar flux dynamics. Inspired by the two classical phenomenological theories of return-to-isotropy (RI) and isotropization-of-production (IP), we assume that the most likely directions of the PS term are around a fan-shaped region bounded by the RI and IP directions. Subsequently, we propose a strategy that requires two additional simulations, defining perturbations of the PS directions towards the RI and IP limits. The approach is applied to simulations of forced heat convection in a complex pin-fin array configuration, and shows favorable monotonic properties and bounding behaviors for various quantities of interest (QoIs) relevant to heat transfer. To conclude, the results are analyzed from the perspective of transverse scalar transport in a shear flow; the analysis indicates that the proposed approach is likely to exhibit monotonic behaviors in a wide range of scalar transport problems.

\end{abstract}


\end{frontmatter}


\section{Introduction}
\label{Section:Introduction}

Reynolds-averaged-Navier-Stokes (RANS) simulations of turbulent flows with passive scalar transport require closure models for the Reynolds stresses and the turbulent scalar fluxes. The epistemic uncertainty in both model forms can affect the accuracy of the results, and tools to assess the corresponding uncertainty in predicted quantities of interest (QoIs) would provide valuable information when using the simulations for engineering design and analysis. Epistemic uncertainty quantification (UQ) of Reynolds stress models has been explored in several recent studies~\cite{Emory11,Gorle12,Iaccarino17,Zeoli18,Gorle19,Duraisamy19}, but quantifying the uncertainty introduced by scalar flux models has received comparatively less attention. 

The most common approach to scalar flux modeling for engineering applications is based on the standard gradient-diffusion hypothesis (SGDH). This hypothesis relates the scalar flux to the product of a scalar diffusion coefficient and the mean scalar gradient. The scalar diffusion coefficient is defined as the ratio of the turbulent viscosity and a turbulent Prandtl number, $\mathrm{Pr}_t$ (or Schmidt number, $\mathrm{Sc}_t$). When investigating the performance of the SGDH, the primary focus has been on the influence of the selected values for $\mathrm{Pr}_t$ or $\mathrm{Sc}_t$. Their considerable effect on the predicted scalar field has been demonstrated in a variety of flow problems, including pollutant dispersion in urban areas~\cite{Riddle04,Gorle10}, reacting flows in propulsion systems~\cite{Baurle04,Gorle13}, particle burning in packed beds~\cite{Frigerio08}, and film cooling in turbomachines~\cite{Ling13,Ling15}. Calibration of \(\mathrm{Pr}_t\) or \(\mathrm{Sc}_t\) has also indicated that it is far from a universal constant; it is highly dependent on the flow configurations and regimes, with a reported range as large as \(0.1\sim2.2\)~\cite{Tominaga07,Combest11}. 

Importantly, studies that focus on the influence of $\mathrm{Pr}_t$ primarily demonstrate the effect of increasing or decreasing the magnitude of the scalar flux vector. They do not investigate the effect of the assumption that the direction of the vector remains identical to the local mean scalar gradient, which has been demonstrated to be incorrect in \textit{a priori} analyses of scalar transport problems. A formal method for UQ of scalar flux models should also address this assumption. A framework to achieve this was first proposed by \citet{Gorle13}. The method used a generalized gradient diffusion hypothesis (GGDH~\cite{Daly70}), which defines a tensorial diffusion coefficient that is a function of the Reynolds stresses, and it propagates perturbations to the Reynolds stresses through the GGDH model to modify the predicted scalar flux. When defining perturbations based on a high-fidelity data set it was shown that this framework could provide a prediction with a confidence interval that encompasses high-fidelity data; this could not be achieved when only modifying $\mathrm{Pr}_t$. In \cite{Hao20b}, this approach was extended to consider the propagation of physics-based perturbations to the Reynolds stresses through different scalar flux models, including a second-moment model that solves three additional transport equations for the scalar fluxes. The results showed that the perturbations provide an interval prediction for the average Nusselt number in a pin-fin heat exchanger that encompasses the value obtained from high-fidelity simulations; however, the intervals predicted for the local distribution of the Nusselt number did not fully encompass these data. It was suggested that this shortcoming is related to uncertainty in the modeled scalar flux.

In this paper, we propose a UQ method that addresses the inherent inadequacies of models for the dynamics (i.e.~the transport equation) of scalar flux. We particularly focus on the pressure scrambling (PS) term, which is the primary mechanism acting to balance the productions. Since the mechanism of PS is not fully understood to date, this term is also a major uncertainty source of second-moment models in practice. Therefore, the basic idea of this work is to `bound' the (vectorial) PS term through a reasonable approach based on our limited understanding of its effects.

\section{Scalar flux model UQ method}
\label{Section:Method}

\subsection{Reviewing the dynamics of scalar flux}
\label{Subsection:Dynamics}

 For incompressible turbulent flow, with constant density \(\rho\), molecular viscosity \(\nu\), and molecular scalar diffusion coefficient \(\kappa\), the transport equation for a mean passive scalar \(\mathit{\Theta}\) is given by:
\begin{equation}
\label{eq:meanscalar}
    \frac{\bar{D}}{\bar{D}t}\mathit{\Theta} = \frac{\partial}{\partial x_i}\left(\kappa\frac{\partial\mathit{\Theta}}{\partial x_i}-\overline{u_i^\prime\theta^\prime}\right),
\end{equation}
where \(U_i\) and \(\mathit{\Theta}\) are the mean velocity and scalar fields, \(u_i^\prime\) and \(\theta^\prime\) the corresponding turbulent fluctuations, and \(\bar{D}/\bar{D}t\equiv\partial/\partial t+U_i\,\partial/\partial_i\) is the material derivative. The term \(\overline{u_i^\prime\theta^\prime}\) in Eq.~(\ref{eq:meanscalar}) is the turbulent scalar flux to be closed. 

The exact transport equation for \(\overline{u_i^\prime\theta^\prime}\) reads:
\begin{equation}
\label{eq:scalarflux}
\begin{aligned}
    \frac{\bar{D}}{\bar{D}t}\overline{u_i^\prime\theta^\prime} = & -\overline{u_i^\prime u_j^\prime} \frac{\partial\mathit{\Theta}}{\partial x_j} - \overline{u_j^\prime\theta^\prime} \frac{\partial U_i}{\partial x_j} + \overline{\frac{p^\prime}{\rho}\frac{\partial\theta^\prime}{\partial x_i}} -\left(\nu+\kappa\right)\overline{\frac{\partial u_i^\prime}{\partial x_j}\frac{\partial\theta^\prime}{\partial x_j}} \\& + \frac{\partial}{\partial x_j}\left(\nu\overline{\theta^\prime\frac{\partial u_i^\prime}{\partial x_j}}+\kappa\overline{u_i^\prime\frac{\partial\theta^\prime}{\partial x_j}}-\overline{u_i^\prime u_j^\prime\theta^\prime}-\overline{\frac{p^\prime}{\rho}\theta^\prime}\delta_{ij}\right).
\end{aligned}
\end{equation}
On the right-hand-side of Eq.~(\ref{eq:scalarflux}), the first two terms, denoted as \(G_i^\mathrm{I}\) and \(G_i^\mathrm{II}\) hereafter, represent the direct interactions between the mean gradient fields and the large scale turbulent eddies. These interactions are immediately responsible for the production of \(\overline{u_i^\prime\theta^\prime}\); they are in closed form given the Reynolds stress field \(\overline{u_i^\prime u_j^\prime}\).
The third term, denoted hereafter as \(\mathit{\Pi}_i\), represents the pressure scrambling (PS) effects. This is the primary mechanism balancing the two production terms, but it is not well understood to date, primarily due to the highly non-local nature of pressure.
The fourth term, denoted hereafter as \(\varepsilon_{\theta i}\), represents the molecular dissipation; it is usually negligible when the Reynolds and P\'eclet numbers are both high enough, as is the case for the flow considered in this paper. The last four terms, denoted together as \(\mathcal{D}_{\theta i}\), are all in divergence form and represent the diffusive transport. Despite their complex form, it seems that employing a certain type of GDH model (to be introduced in \S\ref{Subsection:Perturbation}) for the entirety causes no serious problems in most engineering cases (e.g.~see Sec.~2.4 in~\cite{Gatski96}). As a result, the performance of second-moment models for Eq.~(\ref{eq:scalarflux}) largely depends on the closure for the PS term \(\mathit{\Pi}_i\); this motivates our focus on quantifying the uncertainty in this term.

\subsection{Theoretical basis on the pressure scrambling effects}
\label{Subsection:Classic}
  Existent understanding of the PS effects is informed by the wave nature of the pressure fluctuation \(p^\prime\), which tends to disorganize the turbulent eddy structures and thus decorrelate two variables, in this case \(u_i^\prime\) and \(\theta^\prime\). Furthermore, the governing Poisson equation for \(p^\prime\)
\begin{equation}
\label{eq:poisson}
    -\frac{1}{\rho}\frac{\partial^2 p^\prime}{\partial x_i\partial x_i} = \frac{\partial^2}{\partial x_i\partial x_j}\left(u_i^\prime u_j^\prime - \overline{u_i^\prime u_j^\prime}\right) + 2\frac{\partial U_i}{\partial x_j}\frac{\partial u_j^\prime}{\partial x_i}
\end{equation}
indicates that three separate mechanisms contribute to \(p^\prime\): the slow pressure \(p^{(s)}\) induced by the interaction between turbulent eddies; the rapid pressure \(p^{(r)}\) induced by the interaction between turbulent eddies and mean flow distortion; and the harmonic pressure \(p^{(h)}\) induced by the boundary conditions, e.g. wall-blocking and free-surface reflection. Correspondingly, the wave effects of \(p^\prime=p^{(s)}+p^{(r)}+p^{(h)}\) can also be discussed in terms of these separate mechanisms, and the PS term in Eq.~(\ref{eq:scalarflux}) can be written as \(\mathit{\Pi}_i=\mathit{\Pi}^{(s)}_i+\mathit{\Pi}^{(r)}_i+\mathit{\Pi}^{(h)}_i\). In this analysis, we focus on the uncertainty in the PS term resulting from the turbulent nature of the flow, i.e. on \(\mathit{\Pi}^{(s)}_i\) and \(\mathit{\Pi}^{(r)}_i\). 

First, a theoretical basis for modeling the slow part \(\mathit{\Pi}^{(s)}_i\), is provided by the return-to-isotropy (RI) hypothesis (\citet{Rotta51,Monin65}), which argues that the vector \(\mathit{\Pi}^{(s)}_i\) tends to diminish the scalar flux. This implies that its direction should be opposite the scalar flux direction:
\begin{equation}
\label{eq:RI}
    \mathit{\Pi}^{(s)}_i \quad /\!\!/ \quad -\overline{u_i^\prime\theta^\prime}.
\end{equation}
Second, the theoretical basis for modeling the rapid part \(\mathit{\Pi}^{(r)}_i\), is provided by the isotropization-of-production (IP) hypothesis (\citet{Naot70,Owen74}), which argues that the vector \(\mathit{\Pi}^{(r)}_i\) tends to counteract the scalar flux production by mean flow distortion, i.e.~\(G^\mathrm{II}_i\). This implies that its direction should be opposite the direction of \(G^\mathrm{II}_i\):
\begin{equation}
\label{eq:IP}
    \mathit{\Pi}^{(r)}_i \quad /\!\!/ \quad -G^\mathrm{II}_i.
\end{equation}

Eqs.~(\ref{eq:RI}) and (\ref{eq:IP}) have been used to establish PS models through calibration in canonical flows. For example, a basic model has been proposed based on calibration in homogeneous shear flow with a transverse scalar gradient \cite{Launder75}. The drawback of this approach is that the accuracy of these models can not be guaranteed when extrapolating to more complex engineering flows: the lack of information on the magnitudes of the slow and rapid pressure terms in Eqs.~(\ref{eq:RI}) and (\ref{eq:IP}) makes the model highly dependent on the calibration process. However, from the perspective of model UQ, the limited information provided by the RI and IP hypotheses supports identifying plausible directions of the combined PS term \(\mathit{\Pi}^{(s)}_i+\mathit{\Pi}^{(r)}_i\): directions close to the RI-IP plane are more likely to occur than directions considerably deviating from the plane. This provides the basis for the proposed UQ approach.

\subsection{Baseline model and perturbation strategy}
\label{Subsection:Perturbation}

As a starting point, we adopt the basic second-moment model \cite{Launder75,Launder89} as the baseline (BSL) model for the UQ framework. In this model, the pressure scrambling terms are represented as:
\begin{equation}
\label{eq:basicmodel}
    \mathit{\Pi}^{(s)}_i + \mathit{\Pi}^{(r)}_i =  -c_{\theta1}\frac{\varepsilon}{k}\overline{u_i^\prime\theta^\prime} -c_{\theta2}G^\mathrm{II}_i,
\end{equation}
with $c_{\theta 1}=2.9$ and $c_{\theta 2}=0.4$. The dissipation $\varepsilon_{\theta i}$ is assumed to be negligible, and the last four terms in Eq.~(\ref{eq:scalarflux}) are modeled as:
\begin{equation}
\label{eq:basicmodel2}
\begin{aligned}
 \mathcal{D}_{\theta i} = \frac{\partial}{\partial x_j}\left(c_{\theta d}\frac{k}{\varepsilon}\overline{u_j^\prime u_k^\prime}\frac{\partial\overline{u_i^\prime\theta^\prime}}{\partial x_k}\right)
\end{aligned}
\end{equation}
with $c_{\theta d}=0.15$, \(k\equiv\overline{u_i^\prime u_i^\prime}/2\) the turbulent kinetic energy, and \(\varepsilon\equiv\nu\overline{\left(\partial_j u_i^\prime\right)\left(\partial_j u_i^\prime\right)}\) the energy dissipation rate. 

\begin{figure}[htbp]
\centering\includegraphics[width=0.50\linewidth]{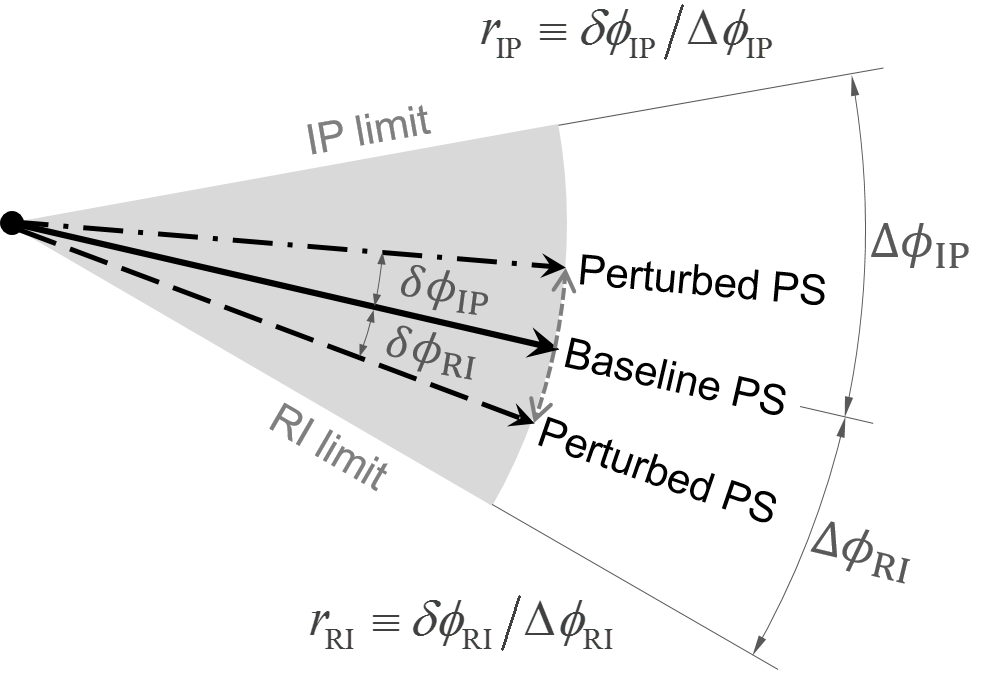}
\caption{The strategy of perturbing the PS direction in the UQ approach}
\label{fig:perturbation}
\end{figure}

Eq.~\ref{eq:basicmodel} implies that the baseline PS term lies on the RI-IP plane. As discussed in Section \ref{Subsection:Classic}, its exact direction on this plane is an important source of model uncertainty, since it is determined by the relative magnitudes of the slow and rapid terms. Hence, we propose a simple perturbation strategy to quantify this uncertainty: for a local PS vector predicted by the BSL model, we perturb its direction within the RI-IP plane as shown in Fig.~\ref{fig:perturbation}. The perturbation requires the definition of the uncertain parameter \(r_\mathrm{RI}\) or \(r_\mathrm{IP}\) in the interval [0,1], where 0 corresponds to no perturbation, while 1 corresponds to a perturbation to the RI or IP limit. In this initial analysis, we prescribe a single value of \(r_\mathrm{RI}\) or \(r_\mathrm{IP}\) and apply it uniformly in the entire computational domain.

\section{Results}
\label{Section:Results}

\subsection{Test case}
\label{Subsection:Setup}

\begin{figure}[htbp]
\centering\includegraphics[width=0.90\linewidth]{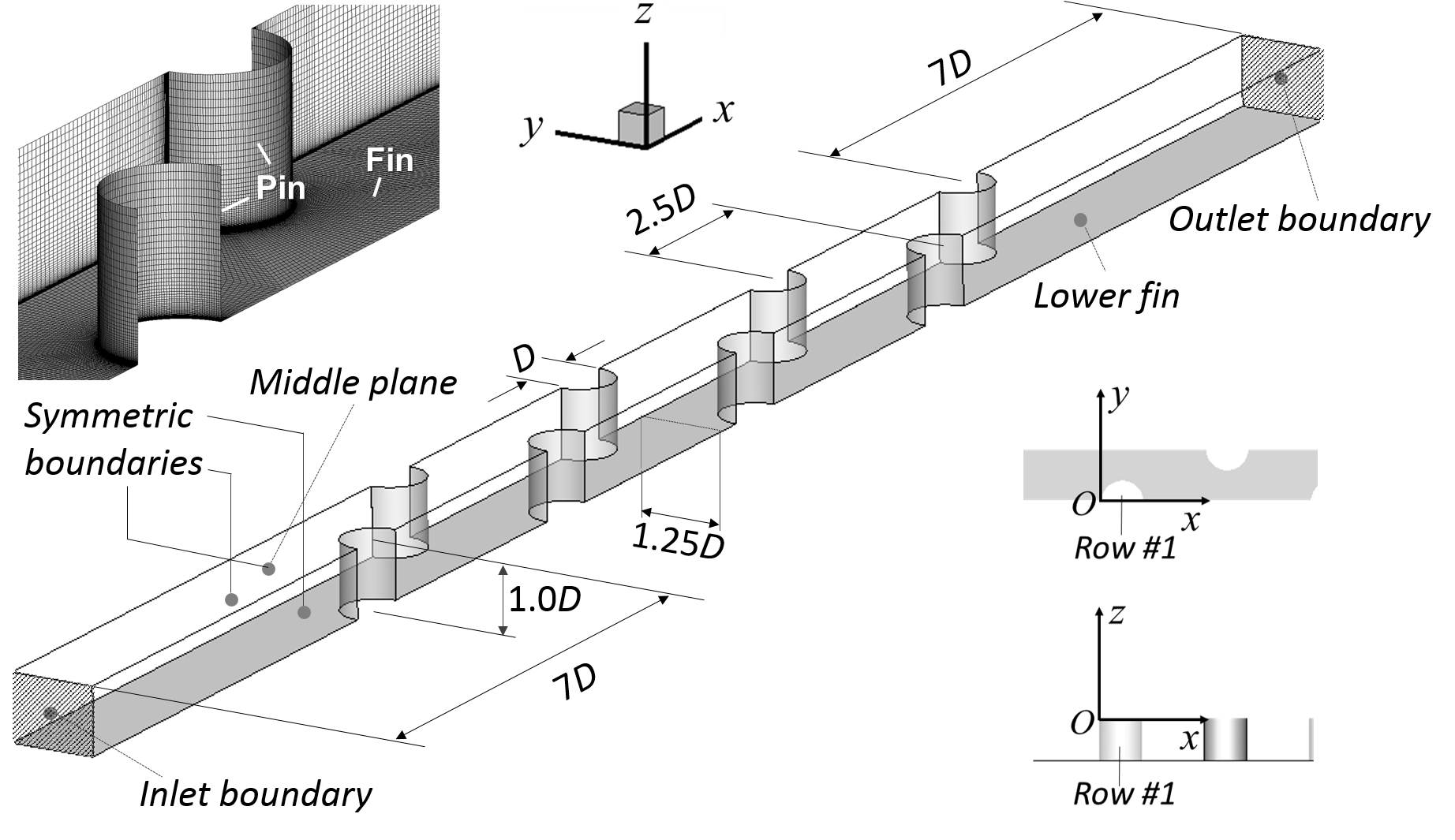}
\caption{Computational domain and mesh for the pin-fin array configuration}
\label{fig:pinfinarray}
\end{figure}

The test case considers heat transfer in a pin-fin array consisting of eight rows of staggered cylinders (pins) between two parallel flat plates (fins). It was studied experimentally by \cite{Ames05,Ames06,Ames07}, and we have reported a high-fidelity LES~\cite{Hao19} for the operating conditions considered in this study. Fig.~\ref{fig:pinfinarray} shows the computational domain, which spans half of the lateral (\(y\)) pin spacing and half of the vertical (\(z\)) fin spacing due to symmetry in both directions. 
All pin and fin surfaces have a fixed and uniform temperature \(\mathit{\Theta}_w\). At the inlet, fluid with a higher temperature \(\mathit{\Theta}_{in}\) enters the channel, inducing forced heat convection between the fluid and the pin and fin surfaces. The bulk Reynolds number is given by \(\mathrm{Re}_D=V_mD/\nu=10^4\), where \(D\) is the diameter of the cylindrical pins and \(V_m\) is the average velocity on the throat cross-section between two laterally adjacent pins. The Prandtl number \(\mathrm{Pr}\) is 0.71. Only steady solutions (i.e.~\(\partial/\partial t = 0\)) of Eqs.~(\ref{eq:meanscalar}) and (\ref{eq:scalarflux}) are considered. Details on the numerical set-up can be found in \cite{Hao20}.

The main QoIs are the heat transfer rates on the pin and fin surfaces, characterized by the Nusselt number \(\mathrm{Nu}\equiv(\partial_n\mathit{\Theta})_wD/(\mathit{\Theta}_b-\mathit{\Theta}_w)\), where \((\partial_n\mathit{\Theta})_w\) is the wall-normal temperature gradient and \(\mathit{\Theta}_b\) is the bulk temperature of the surrounding fluid.

\subsection{Analysis of PS directions using the database}
\label{Subsection:Validity}

The PS term \(\overline{(p^\prime/\rho)\partial_i\theta^\prime}\) cannot be directly obtained from the LES since the contribution from unresolved fluctuations cannot be neglected. Hence, we estimate the term through the budget of Eq.~(\ref{eq:scalarflux}), with \(\mathcal{D}_{\theta i}\) and \(\varepsilon_{\theta i}\) modeled as in the BSL model:
\begin{equation}
\label{eq:psbudget}
\begin{aligned}
    \overline{\frac{p^\prime}{\rho}\frac{\partial\theta^\prime}{\partial x_i}} \approx U_j\frac{\partial\overline{u_i^\prime\theta^\prime}}{\partial x_j} + \overline{u_i^\prime u_j^\prime} \frac{\partial\mathit{\Theta}}{\partial x_j} + \overline{u_j^\prime\theta^\prime} \frac{\partial U_i}{\partial x_j} - \frac{\partial}{\partial x_j}\left(c_{\theta d}\frac{k}{\varepsilon}\overline{u_j^\prime u_k^\prime}\frac{\partial\overline{u_i^\prime\theta^\prime}}{\partial x_k}\right)
\end{aligned}
\end{equation}
The fields \(U_i\), \(\mathit{\Theta}\), \(k\), \(\overline{u_i^\prime u_j^\prime}\), and \(\overline{u_i^\prime\theta^\prime}\) are all directly obtained from the LES data, while \(\varepsilon\) is partially modeled as in \cite{Hao19}.

\begin{figure}[htbp]
\centering\includegraphics[width=0.64\linewidth]{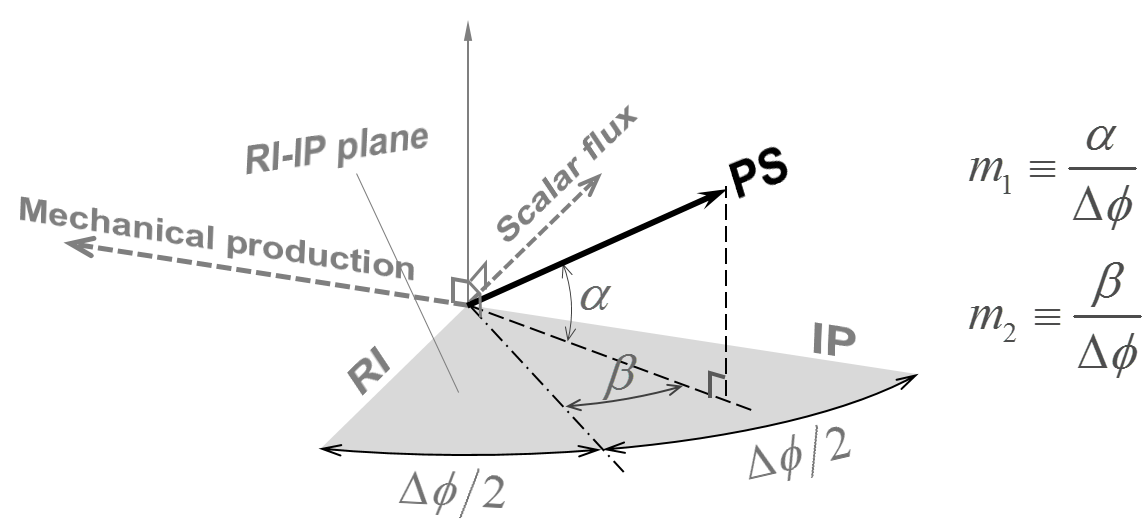}
\caption{Definition of the metrics \(m_1\) and \(m_2\), indicating the relation between the estimated PS direction and the RI (\(m_1=0\), \(m_2=-1/2\)) and IP (\(m_1=0\), \(m_2=+1/2\)) directions.}
\label{fig:metrics}
\end{figure}

To analyze the direction of the resulting PS vectors, we define two metrics \(m_1\) and \(m_2\) as shown in Fig.~\ref{fig:metrics}; these metrics quantify the discrepancy between the estimated PS direction and the RI and IP directions.

\begin{figure}[htbp]
\centering\includegraphics[width=0.5\linewidth]{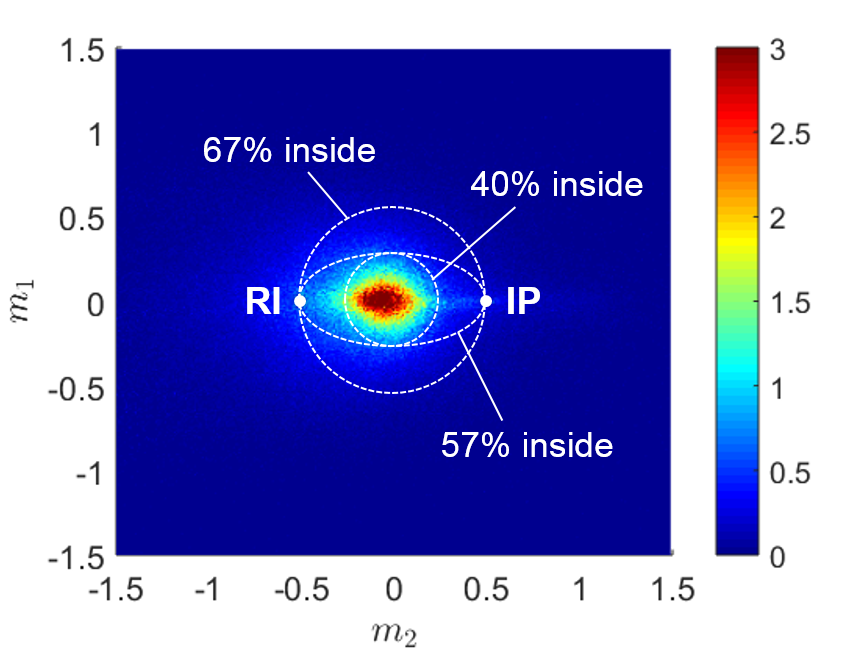}
\caption{Joint distribution of \((m_1,m_2)\) to visualize the estimated PS directions, weighted by their magnitude, relative to the RI and IP directions over the entire test section of the pin-fin array.}
\label{fig:pinfinvald}
\end{figure}

Fig.~\ref{fig:pinfinvald} shows the joint distribution of $m_1$ and $m_2$, weighted by the PS term magnitude, over the entire domain. The most likely PS directions concentrate around the RI-IP plane, and 57\% of the values are inside the ellipse \(\sqrt{4m_1^2+m_2^2}=1/2\). This result confirms the validity of the hypothesis that directions close to the RI-IP plane are more likely to occur than directions considerably deviating from the plane.

\subsection{Practical performance of the UQ method}
\label{Subsection:Monotonicity}

In this section we present results obtained from a series of simulations with different \(r_\mathrm{RI}\) or \(r_\mathrm{IP}\) values. To evaluate the performance of the approach in absence of Reynolds stress model errors, the simulations solve Eqs.~(\ref{eq:meanscalar}) and ~(\ref{eq:scalarflux}) using the BSL model with input for \(U_i\), \(k\), \(\varepsilon\) and \(\overline{u_i^\prime u_j^\prime}\) obtained from the LES data.

\begin{figure}[htbp]
\centering\includegraphics[width=0.47\linewidth]{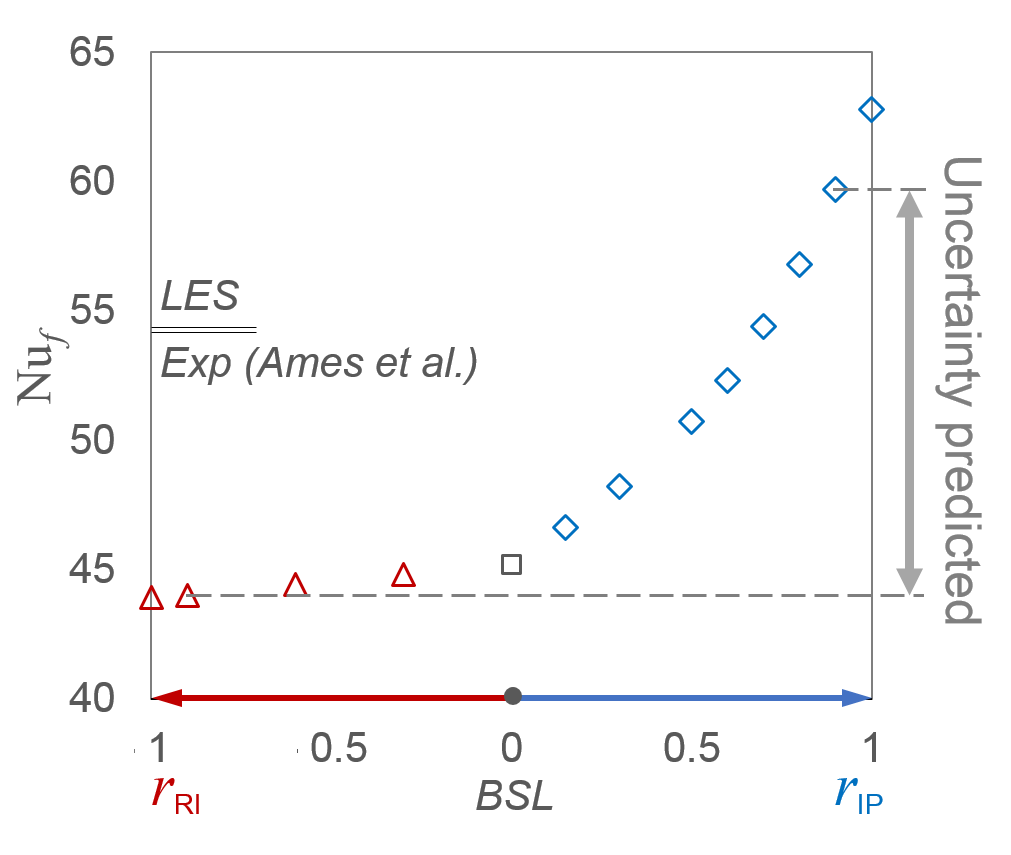}
\caption{Globally averaged Nusselt numbers on the fins changed with the uncertain parameter}
\label{fig:mono}
\end{figure}

\begin{figure}[htbp]
\centering\includegraphics[width=0.8\linewidth]{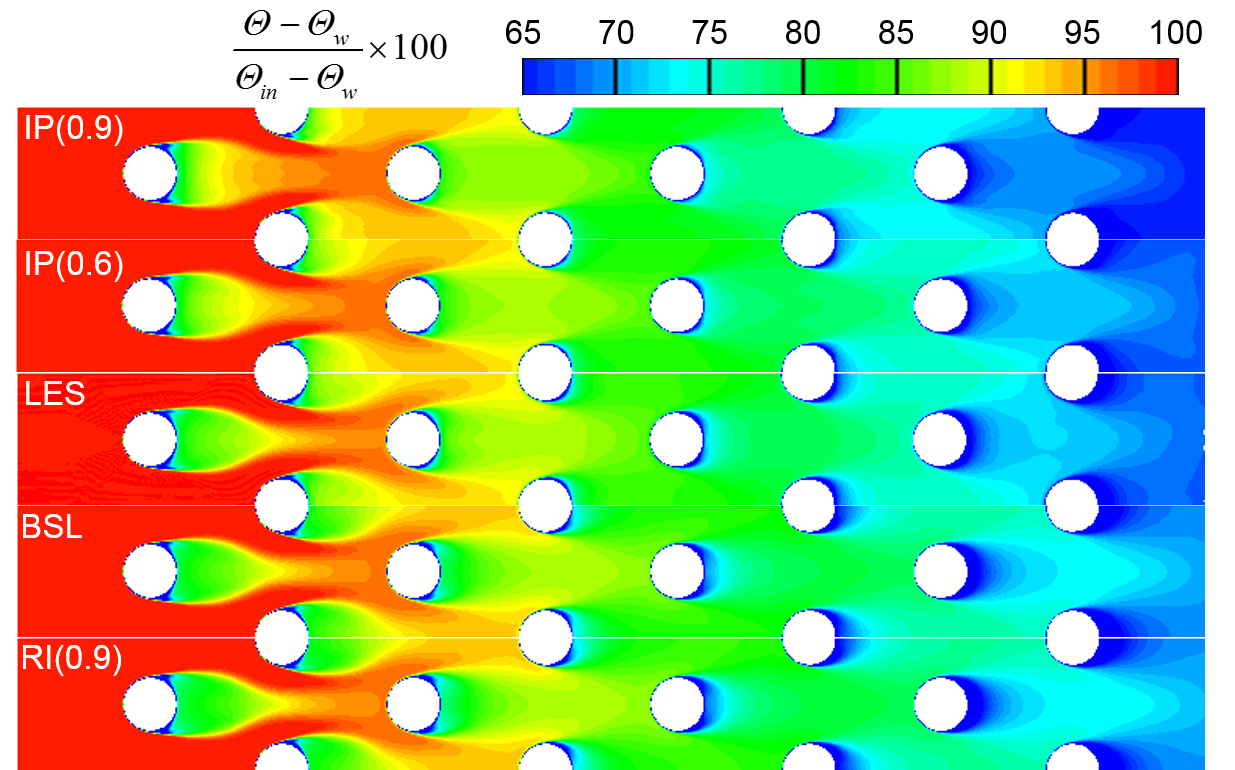}
\caption{Temperature counters on the middle plane of the pin-fin array}
\label{fig:temperature}
\end{figure}

First, Fig.~\ref{fig:mono} shows the globally averaged Nusselt numbers on the fins, \(\mathrm{Nu}_f\) for the full range of the uncertain parameter \(r_\mathrm{IP}\) or \(r_\mathrm{RI}\). With \(r_\mathrm{IP}\) changing from 0 to 1, \(\mathrm{Nu}_f\) monotonically increases by up to 38\%; in contrast, with \(r_\mathrm{RI}\) increasing, \(\mathrm{Nu}_f\) monotonically decreases, to a far lesser extent, by up to 3\%. A similar trend is reflected by the temperature contours in Fig.~\ref{fig:temperature}. From the bottom to the top, the fluid cooling is gradually enhanced, as indicated by decreasing outlet temperatures.
Given the monotonicity, and the observation that the interval obtained from simulations with \(r_\mathrm{RI,\,IP}=0.9\) encompasses the LES and experimental results, these results will be used to define the predicted uncertainty interval for the local Nusselt numbers in the remainder of this section.

\begin{figure}[htbp]
\centering\includegraphics[width=0.49\linewidth]{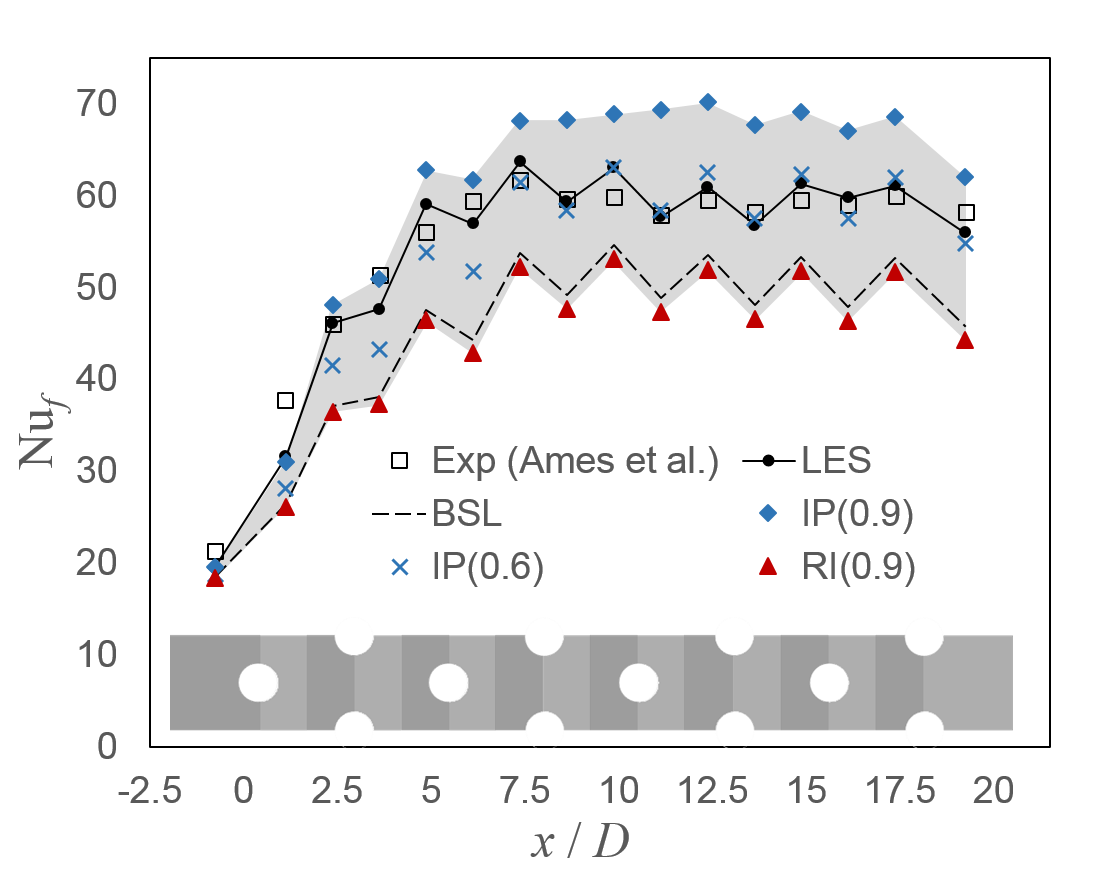} \includegraphics[width=0.49\linewidth]{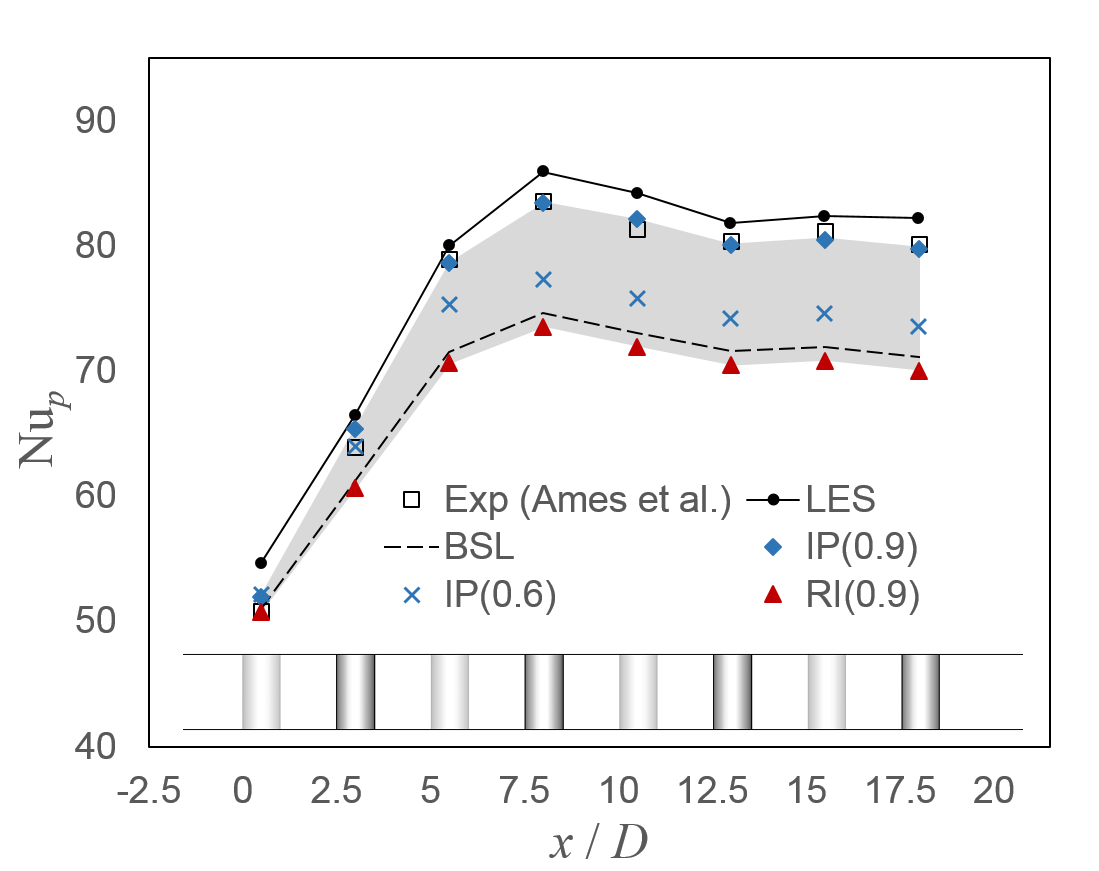}
\caption{Locally averaged Nusselt numbers on the fins (left) and the pins (right)}
\label{fig:nuf}
\end{figure}

Fig.~\ref{fig:nuf} shows two local Nusselt numbers: \(\mathrm{Nu}_f\) averaged over individual \(1.25D\times2.5D\) small fin patches on the left, and \(\mathrm{Nu}_p\) averaged over the surfaces of individual pins on the right. Compared to the LES results, the BSL model underestimates the local values by \(14\sim21\%\) for \(\mathrm{Nu}_f\), and by \(11\sim14\%\) for \(\mathrm{Nu}_p\). The perturbation with \(r_\mathrm{IP}=0.9\) substantially raises \(\mathrm{Nu}_f\), reaching a level \(8\sim17\%\) higher than the LES results. Conversely, the perturbation with \(r_\mathrm{RI}=0.9\) slightly reduces the \(\mathrm{Nu}_f\) prediction to a level \(16\sim23\%\) lower than LES. The resulting interval encompasses the LES predictions for \(\mathrm{Nu}_f\) in the entire test section. The results for $\mathrm{Nu}_p$ exhibit a similar trend, altough the \(r_\mathrm{IP}=0.9\) perturbation still slightly underpredicts the LES values by \(1\sim3\%\).

\begin{figure}[htbp]
\centering
\includegraphics[width=0.5\linewidth]{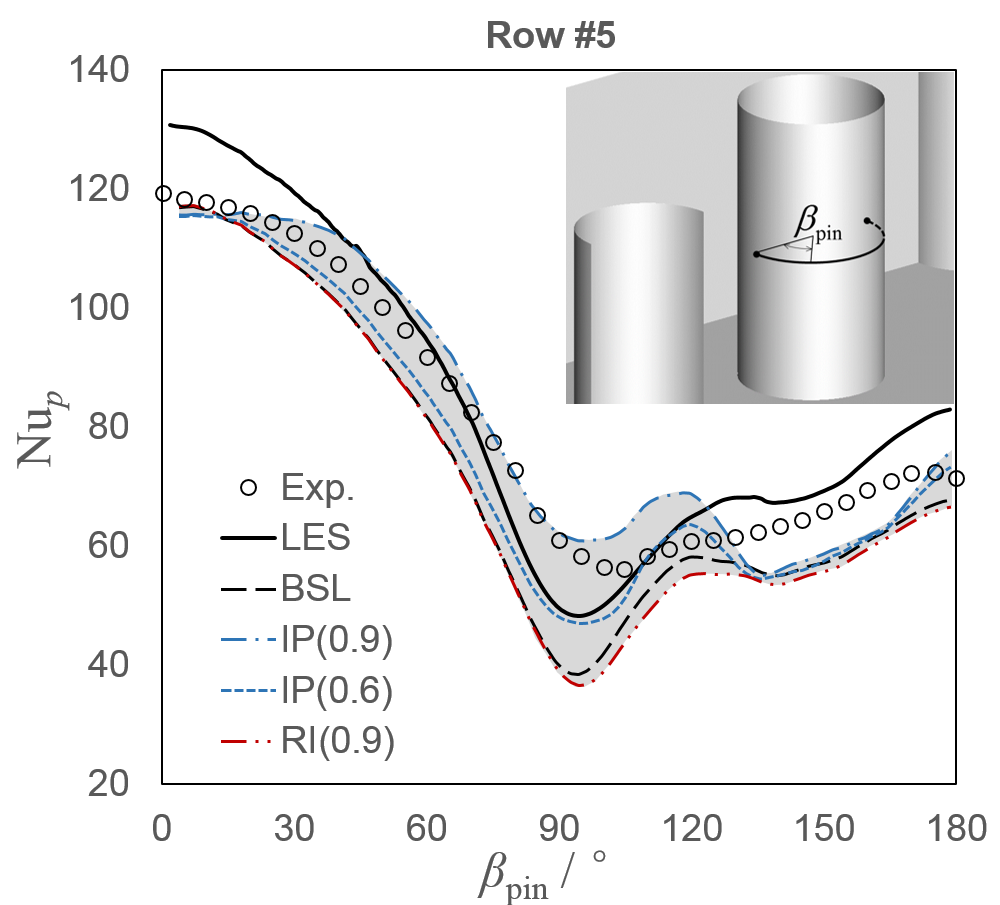}
\caption{Nusselt number distributions on the pin surface (row \#5)}
\label{fig:nup_row5}
\end{figure}

Fig.~\ref{fig:nup_row5} depicts the local distribution of the Nusselt number along the pin on row \# 5. The plot indicates that the small under prediction of $\mathrm{Nu}_p$ in Fig.~\ref{fig:nuf} is due to the small uncertainty predicted around the stagnation point and the downstream wake region. The narrow uncertainty in the prediction for the upwind values  (\(0^\circ\leq\beta_\mathrm{pin}<35^\circ\)) results from small angles between the RI and IP directions in these regions: due to symmetry the angle is zero at \(\beta_\mathrm{pin}=0^\circ\), and there is no perturbation to the PS directions. In the downstream region (\(135^\circ<\beta_\mathrm{pin}\leq180^\circ\)), the result indicates that the perturbations do not strongly affect the heat flux from the bulk flow to the wake region; this deficiency will be a focus of future work.

\section{Further discussions}
\label{Section:Discussions}

The results in Section~\ref{Section:Results} clearly demonstrate the monotonicity of QoIs (Nusselt numbers) with respect to \(r_\mathrm{RI}\) or \(r_\mathrm{IP}\), a favorable property for an interval UQ approach. In this section, we preliminarily analyze the reason for this characteristic from one perspective.

\begin{figure}[htbp]
\centering
\includegraphics[width=0.4\linewidth]{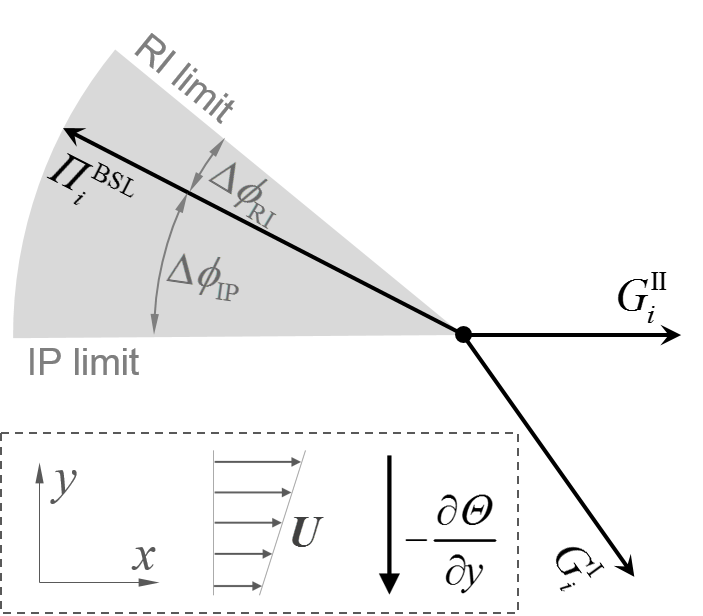}
\caption{Sketch of a typical scalar flux budget in a parallel shear flow with \(\partial U_x/\partial y>0\) and \(\partial\mathit{\Theta}/\partial y>0\)}
\label{fig:budget}
\end{figure}

Figure ~\ref{fig:budget} depicts the case of a parallel shear flow:  \(U_x>0\) and \(\partial U_x/\partial y>0\) are the only non-zero components of the velocity and the velocity gradient, and we assume a mean scalar gradient \(\partial\mathit{\Theta}/\partial y>0\) in the transverse (\(y\)) direction. If the turbulence is close to a state of local equilibrium, we have \(\overline{u_x^\prime u_y^\prime}<0\), \(\overline{u_y^\prime\theta^\prime}<0\) and \(\overline{u_x^\prime\theta^\prime}>0\); consequently, the vectors of the two production terms \(G^\mathrm{I,\,II}_i\) and the BSL predicted PS term \(\mathit{\Pi}^\mathrm{BSL}_i\) can be illustrated as shown in Fig.~\ref{fig:budget}. Given these relative directions, a perturbation of the PS direction towards the IP limit will increase the magnitude of \(\overline{u_y^\prime\theta^\prime}\), while a perturbation towards the RI limit will decrease the magnitude of \(\overline{u_y^\prime\theta^\prime}\). Since this is the flux component directly responsible for the transverse scalar transport, and the most relevant component to the QoIs in general, this explains the monotonic behavior of the predictions as a function of \(r_\mathrm{RI}\) or \(r_\mathrm{IP}\).

Additionally, we note that the flux directions predicted by \(\mathit{\Pi}^\mathrm{BSL}_i\) are usually closer to the RI than the IP direction, which causes the higher sensitivity of the QoIs to \(r_\mathrm{IP}\) than to \(r_\mathrm{RI}\). This asymmetry is consistent with results in a parallel shear flow: available experimental data for homogeneous shear turbulence~\cite{Tavoularis81} indicates that the ratio of  \(\Delta\phi_\mathrm{RI}\) and \(\Delta\phi_\mathrm{IP}\) is approximately \(1/3\sim1/2\).

\section{Conclusions}
\label{Section:Conclusions}

This paper proposes an approach to quantify scalar flux model uncertainties in RANS simulations of turbulent scalar transport. The approach addresses the inherent inadequacy of the model for the pressure scrambling (PS) term in the transport equations of scalar flux. Specifically, it perturbs the PS directions predicted by the basic second-moment model within a fan-shaped region bounded by the directions corresponding to the return-to-isotropy (RI) and isotropization-of-production (IP) theories.

The UQ method is applied to the prediction of heat transfer in a pin-fin array. First, high-fidelity LES data are analyzed to support the proposed perturbation approach: PS directions estimated through the scalar flux budget are shown to concentrate around the RI-IP plane. Next, the results demonstrate that the approach exhibits favorable monotonic behavior: perturbations of the PS directions towards the IP (or RI) limit consistently enhance (or suppress) heat transfer. The resulting plausible intervals for the QoIs, provided by the results of \(r_\mathrm{IP,\,RI}=0.9\), encompass the LES predictions for the globally averaged and the locally averaged fin Nusselt numbers. Finally, we explain the monotonic behavior of the proposed approach from the perspective of transverse scalar transport in shear flows, which indicates that the proposed UQ approach is likely to exhibit monotonic behavior in a wide range of scalar transport problems.

To further improve the method and promote its use in engineering applications we identify two future areas of research. First, the analysis of the LES data indicates that some PS directions are out of the RI-IP plane. As shown in Section \ref{Section:Discussions}, perturbations out of this plane seem unnecessary for estimating plausible intervals for the QoIs in simple cases, but this should be further investigated for more complex flows. Second, the current method does not introduce uncertainty in the magnitude of the PS vector. The need to introduce this additional uncertainty should be further investigated using analysis of high-fidelity data, or testing of other established PS models (see \cite{Hanjalic11} for review). To conclude, it is worth noting that the proposed method provides a basis for exploring data-driven approaches where high-fidelity data could be used to more accurately inform the PS vector perturbations and reduce the uncertainty in the predictions.

\section{Acknowledgements}
The research was funded by EUFORIA (grant number IWT-140068).

\bibliographystyle{model1-num-names}
\bibliography{sample.bib}

\end{document}